\documentclass[conference, 10pt]{IEEEtran}

\usepackage[T1]{fontenc}
\usepackage{tikz}
\usepackage{amsmath}
\usepackage{subcaption}
\usepackage{tcolorbox}
\usepackage{graphicx}

\tcbuselibrary{skins}

\title{\LARGE \bf
Finding (and exploiting) vulnerabilities on IP Cameras: the Tenda CP3 case study}

\author{%
  \IEEEauthorblockN{%
  Dario Stabili\IEEEauthorrefmark{1},
  Tobia Bocchi\IEEEauthorrefmark{2},
  Filip Valgimigli\IEEEauthorrefmark{2} and
  Mirco Marchetti\IEEEauthorrefmark{2}
  }%
  \IEEEauthorblockA{\IEEEauthorrefmark{1} Alma Mater Studiorum - University of Bologna
  \\Department of Computer Science and Engineering
  \\dario.stabili@unibo.it}%
  \IEEEauthorblockA{\IEEEauthorrefmark{2} University of Modena and Reggio Emilia
  \\Department of Engineering ``Enzo Ferrari''
  \\\{tobia.bocchi, filip.valgimigli, mirco.marchetti\}@unimore.it}%
}

\begin{document}






\maketitle

\begin{abstract}
Consumer IP cameras are now the most widely adopted solution for remote monitoring in various contexts, such as private homes or small offices. While the security of these devices has been scrutinized, most approaches are limited to relatively shallow network-based analyses. In this paper, we discuss a methodology for the security analysis and identification of remotely exploitable vulnerabilities in IP cameras, which includes static and dynamic analyses of executables extracted from IP camera firmware. Compared to existing methodologies, our approach leverages the context of the target device to focus on the identification of malicious invocation sequences that could lead to exploitable vulnerabilities.
We demonstrate the application of our methodology by using the Tenda CP3 IP camera as a case study. We identified five novel CVEs, with CVSS scores ranging from $7.5$ to $9.8$. To partially automate our analysis, we also developed a custom tool based on \emph{Ghidra} and \emph{rhabdomancer}. 


\end{abstract}

\section{Introduction}
\label{s:introduction}

This paper discusses the application of a classical methodology for conducting in-depth security analysis with a focus on consumer IP cameras. IP cameras are popular IoT devices, and their cybersecurity has been scrutinized by the scientific community in recent years. Related works~\cite{li2023let,Li2020Privacy} have already identified several vulnerabilities, but their analysis has mainly focused on network traffic. The discussed methodology takes an in-depth approach and requires physical disassembly of the device. While this attacker model may seem more powerful than the one used in related literature, we argue that it is more realistic and allows for the identification of a higher number of more relevant software vulnerabilities. Consumer IP cameras are inexpensive, making it easy for attackers to purchase one or more devices to experiment on. Moreover, our results demonstrate that this kind of analysis enables attackers to extract sensitive information and devise attack strategies that can be weaponized against other IP cameras of the same make and model, even without physical access. The adopted methodology comprises five main steps. 

The first step involves gathering information from open and public sources related to the IP camera under analysis. This step does not require the camera itself and can be performed by checking public data records from the \emph{FCC ID Search} web service~\cite{fccid}. This information proves to be extremely useful, as it often includes high-resolution pictures of the printed circuit boards, which can be used to identify diagnostic and programming interfaces. Additionally, valuable information can be obtained from firmware repositories. Depending on the camera manufacturer and model, it is often possible to download a copy of the firmware from official or unofficial repositories. This step may enable an attacker to focus on a limited number of IP cameras that are more likely to exhibit interesting vulnerabilities.

The second step requires physical access to a specimen of the IP camera under analysis. Within this step, the camera is disassembled to gain physical access to the inner printed circuit board. The primary objective is to visually inspect it and identify one or more vulnerability surfaces that enable direct and low-level interaction with the camera. In our experience, it is highly likely to identify internal USB ports that have no external connectors, as well as simpler debugging and diagnostic interfaces based on standard JTAG or UART protocols. Having physical access allows us to connect and probe these interfaces to confirm that they are active and explore the related attack surfaces. At this step, it is usually possible to interact with the bootloader and potentially gain access to a command-line interpreter as a privileged user.

The third step begins by extracting the firmware deployed on the IP camera. This task can be accomplished either by exploiting low-level read access to the memory of the IP camera through a diagnostic interface or by physically connecting an external reader to the memory chip soldered to the PCB. In some cases, a chip-off might be necessary, although it has not been required in our experience. The extracted firmware is then subjected to common static analysis procedures aimed at identifying relevant partitions, configuration files, scripts, executable files, and cryptographic material.

The fourth step complements the static analysis of the firmware with a dynamic analysis of the network behavior of the IP camera under test. The peculiarity of our approach is that, instead of applying a general approach for static analysis we focus on the main services exposed on the network by the device, thus considering its use-case scenario. By considering the classical usage scenario of the target device, we demonstrate how existing classical tools for static analysis can be tweaked to quickly identify vulnerabilities exposed by the target device, thus preventing the security researcher to manually analyze different potential sources of vulnerabilities that wouldn't lead to exploits.

The fifth and final step builds upon the information gathered in the previous steps to perform a detailed reverse-engineering process of all executables that implement services available from the network. To partially automate this complex step, we developed a novel tool based on \emph{Ghidra}~\cite{ghidra} and \emph{rhabdomancer}~\cite{rhabdomancer} that identifies the functions responsible for handling data received from network connections, builds the invocation sequence list and correctly identifies the thread responsible for each invocation sequence. This allows security researchers to quickly analyze the whole function call sequence of a target network handler to identify potential vulnerabilities in its code, thus helping them to demonstrate potential exploits and identify possible mitigations.

In this paper, we demonstrate the effectiveness of the proposed methodology by focusing on a popular consumer IP camera, the Tenda CP3, as a use case. Our analysis has already led to the publication of five new CVEs, two of which have a CVSS score of $7.5$ and the remaining three have a CVSS score of $9.8$.

\subsection{Related work}
\label{ss:related}
The scientific literature already contains several research papers focusing on the cybersecurity of IoT devices, IP cameras, and video surveillance systems. However, most of this work limits itself to network-based security analysis, either by sniffing and analyzing network traffic or by interacting with exposed network services. Notable contributions in this field include the analysis of the attack surface of IP-based surveillance systems~\cite{Kalbo2020IPBasedSecurity}, as well as wide-ranging analysis of IoT devices~\cite{li2023let} and the exposed network services of IP cameras belonging to a given nation~\cite{Batich2021Exploiting}. These papers demonstrate that connected IoT devices present many vulnerabilities; however, in many cases, the analysis is limited to relatively simple scanning and probing activities.

Other papers more related to our proposal focus on the in-depth analysis of network communications of a given IP camera \cite{Abdalla2020IoTSecurity,Biondi2021VulnerabilityAssessment}. In particular, the authors of~\cite{Biondi2021VulnerabilityAssessment} managed to exploit network-based attacks, such as Man-in-the-Middle, to eavesdrop on and interact with network communications between the IP camera (a TP-Link Tapo C200) and other devices on the same local network, thus identifying three novel vulnerabilities. We remark that these papers only based their analysis on network interactions, without identifying relevant vulnerabilities (such as remote code execution) that require an in-depth reversing of executables extracted from the firmware of IP cameras.

In relevant related work, Shwartz et al.~\cite{shwartz2018pandora} perform firmware extraction from $16$ different IoT devices, with the final goal of extracting and cracking passwords that would allow remote access to the compromised devices. They demonstrate the effectiveness of their approach, as well as the widespread vulnerability related to password management of IoT devices, by creating a modified version of the Mirai botnet that utilized these passwords to compromise vulnerable devices. The authors discuss the possibility of conducting a more comprehensive static analysis and reverse engineering of the analyzed devices but do not undertake these tasks themselves.

On the other hand, this paper introduces a methodology for the security analysis of consumer IP cameras, demonstrated through the examination of the Tenda CP3 as a case study, and the development of a novel tool. To our knowledge, this is the first paper to delve into such an in-depth analysis of IP cameras.

\subsection{Outline}
\label{ss:outline}
The remainder of this paper is organized as follows. Section~\ref{s:tenda_cp3_analysis} presents the IP camera that we use to demonstrate the application of our analysis methodology, while Section~\ref{s:fs_detailed_analysis} presents the detailed analysis of the firmware extracted from our device. Section~\ref{s:binal} analyses and describes the two main programs responsible to handle all external connections to the camera, demonstrating a practical method to identify potential vulnerability and to design their related exploit. Finally, Section~\ref{s:conclusions} highlights the main strengths of this work and describes future development.

\section{Analysis of the Tenda CP3 IP Camera}
\label{s:tenda_cp3_analysis}

In this section, we present the analysis of the Tenda CP3 connected camera and provide details of the hardware platform. Although we had full access to the IP camera during our analysis, our primary focus was on outlining the steps necessary for gathering as much information as possible using OSINT sources and firmware analysis. Consequently, we describe a methodology that could be applied not only to other IP cameras but also to various IoT devices.

\subsection{Hardware analysis}
\label{ss:hardware_analysis}
\begin{figure}[hpbt]
    \centering 
    \begin{subfigure}{.48\columnwidth}
        \centering
        \includegraphics[width=\textwidth]{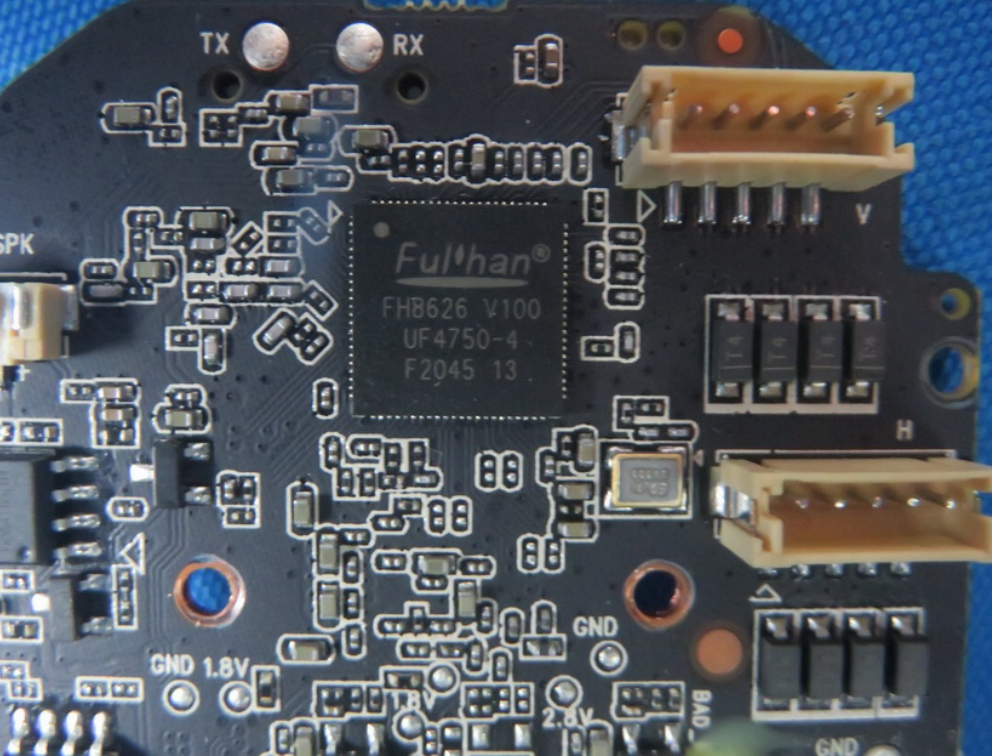}
        \caption{Top side of the main PCB.}
        \label{fig:fccid_internals_front}
    \end{subfigure}
    \hfill
    \begin{subfigure}{.48\columnwidth}
        \centering
        \includegraphics[width=\textwidth]{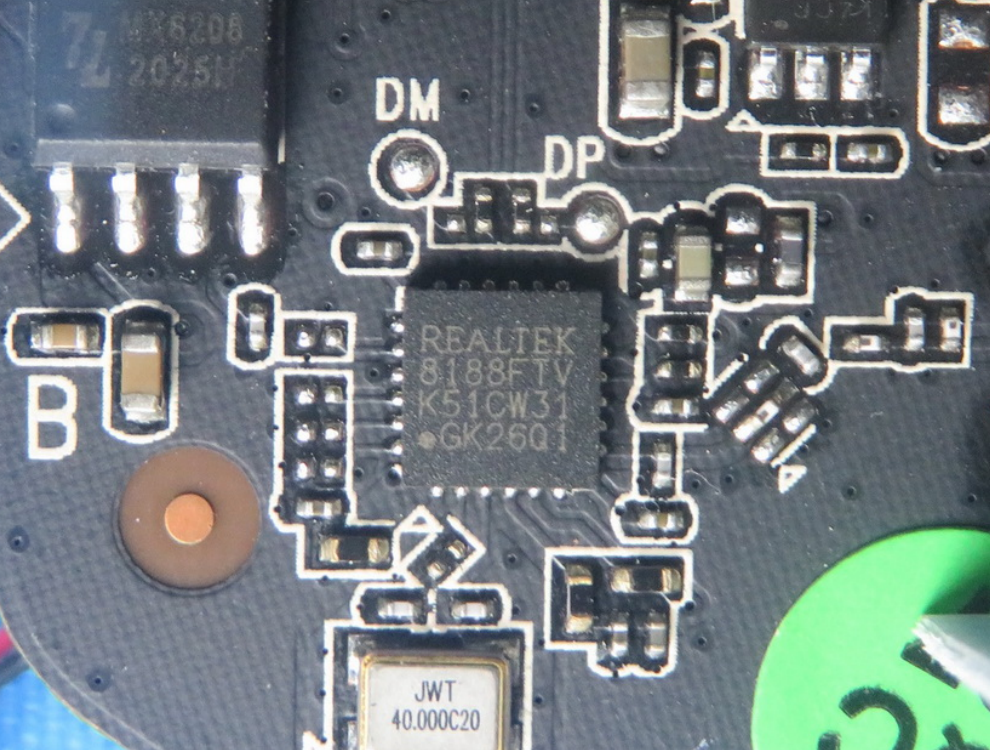}
        \caption{Bottom side of the main PCB.}
        \label{fig:fccid_internals_back}
    \end{subfigure}
    \caption{Internal pictures of the Tenda CP3 accessed via the FCC ID public repository.}
    \label{fig:fccid_internals}
\end{figure}

The Tenda CP3 connected camera is based on a single PCB platform to which the lens and all extension PCBs are connected. We conducted an analysis of the camera's internals using its FCC ID, a unique identifier required for devices transmitting over radio frequencies in the United States. Upon locating the FCC ID on one of the labels attached to the camera, we entered its value into the FCC ID search form available on the Federal Communications Commission website.
The FCC ID of the Tenda CP3 cameras is \texttt{V7TCP3}, with \texttt{V7T} representing the grantee code (i.e., Shenzen Tenda Technology Co., Ltd.) and \texttt{CP3} being the product code assigned by the grantee. By examining the internal pictures of the \texttt{V7TCP3} product (Figure~\ref{fig:fccid_internals}) we identified the main components of the main PCB.

The Tenda CP3 is equipped with a Fullhan \texttt{FH8626 V100} System on Chip designed for HD IP cameras, capable of multi-stream encoding in \textit{H.264} format at 1080p resolution and 15 frames per second. Additionally, it features a Realtek \emph{8188FTV} Network Interface Controller, providing support for \emph{802.11b/g/n 2.4GHz} connectivity.
On the top side of the main PCB, we also identified a UART serial interface, indicated by the two \textit{tx} and \textit{rx} pads located at the top of the image, as well as a flash chip (partially visible on the left side of the picture).

\subsection{UART serial}
\label{ss:uart_serial}
After the initial hardware analysis, we decided to access the internals of the camera available to us to verify the information found via OSINT. Although the labels \emph{tx} and \emph{rx} were not printed on the PCB found inside our device, all the other components and their locations were exactly the same. We proceeded by connecting to the UART serial interface using a USB TTL adapter and minicom configured with a baud rate of $115200$ (8N1).
 
During the first boot of the device we only connected to the \emph{tx} pad of the UART interface with the \emph{rx} pin of our USB TTL adapter to log the output of the system boot process. After analyzing the recorded log of the boot process, we identified several useful information about the bootloader (U-Boot 2010.06-dirty, with the possibility to interrupt the autoboot process by pressing \texttt{`E'}), the OS (a Linux-3.0.8 ARMv7 Linux Kernel Image), the number and mapping of the partitions on the \emph{spi\_flash}, the name of some demons and applications started by the system (one of them being \emph{telenetd}), and some configurations saved by the device, including the configured WiFi credentials printed in clear. 

We then connected the \emph{tx} pin of our USB TTL adapter to the \emph{rx} pad of the PCB to interrupt the autoboot sequence, only to encounter a password-protected login prompt. Later, we will discuss how the password was recovered to access the U-Boot console.

\begin{tcolorbox}[width=.98\columnwidth,bicolor,
    left=0mm,right=0mm,top=0mm,bottom=0mm,arc=0mm,
    colback=gray!10!white,colbacklower=black,colframe=black,
    colupper=black,collower=white]
    \textbf{CVE-2023-30354:} Physical access and WiFi credentials disclosure - Tenda IP Camera CP3 does not defend against physical access to U-Boot via the UART; the Wi-Fi password is shown, and the hard coded boot password can be inserted for console access. 
    \tcblower
    Base Score: \textbf{9.8 Critical} \\
    \scriptsize{Vector: CVSS:3.1/AV:N/AC:L/PR:N/UI:N/S:U/C:H/I:H/A:H}
\end{tcolorbox}
  
\subsection{Firmware extraction and analysis}
\label{ss:firmware_extana}
\begin{figure}[hpbt]
    \centering 
    \includegraphics[width=.50\columnwidth]{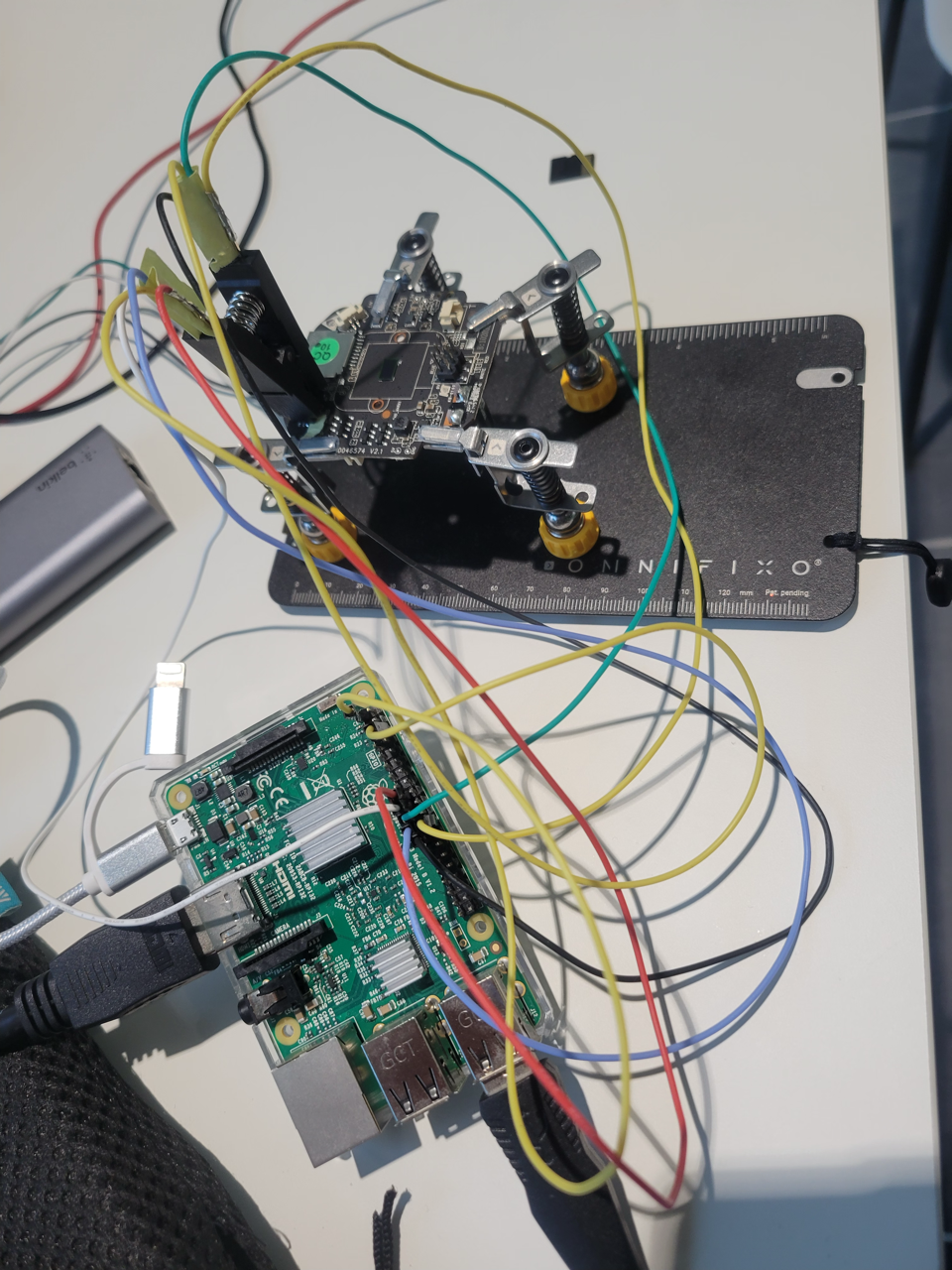}
    \caption{Firmware extraction procedure.}
    \label{fig:flash_chip}
\end{figure}

We extracted the firmware from the SOIC8 chip by directly connecting to the pins of the chip via a dedicated clip, as depicted in Figure~\ref{fig:flash_chip}. We took care to isolate the processor from the flash memory to prevent any modifications during the firmware extraction procedure, thus enabling us to work with the corresponding image of the data found on the flash memory.

We extracted the firmware image using the flashrom utility~\cite{flashrom}, and we modified the permissions on the saved image to be \textit{read-only}. Additionally, we computed a fresh checksum of the saved flash image to ensure that we always had a clean image available.

After obtaining the firmware image, we manually extracted the partitions described in the boot process, resulting in $6$ different files (one for each partition). Here, we provide a high-level analysis of the content of the partitions extracted from the flash image:
\begin{itemize}
    \item \textbf{bootstrap} and \textbf{uboot-env}: partitions containing the configuration of the bootstrap and uboot environment;
    \item \textbf{uboot}: partition containing the uboot bootloader of the device;
    \item \textbf{kernel}: partition containing the root file system of the device. The root file system is stored in a compressed format.
    \item \textbf{data}: partition containing the user data used by the applications being executed on the device. This partition is formatted in a \emph{jff2} file system.
    \item \textbf{app}: partition containing the file system on which all scripts, applications and configuration file of the Tenda CP3 IP camera. This partition is formatted in a \emph{squashfs} file system
\end{itemize}

\section{Detailed analysis of the Tenda CP3 system}
\label{s:fs_detailed_analysis}

The detailed analysis of the Tenda CP3 camera combines both dynamic and static analysis on the content of the \emph{app} partition of the extracted firmware image. Specifically, we are primarily interested in mapping all services being executed on the camera that accept any incoming traffic, and in enumerating interesting files found in the \emph{app} partition of the camera via the firmwalker utility~\cite{firmwalker}. These files include UNIX configuration files (such as \texttt{/etc/passwd} and \texttt{/etc/shadow} files), bash scripts, and all executable files found in the partition.

In this section, we present the results of the analysis of the UNIX configuration files, the service mapping, and the bash scripts responsible for the initialization of the system. In Section~\ref{s:binal}, we provide a detailed analysis of the main programs found active on the camera.

The scripts used in the service and file enumeration process will be publicly released~\footnote[1]{A link to the GitHub repository will be included in the camera-ready version of the manuscript.}.

\subsection{UNIX configuration files}
\label{ss:unix_conf}
We identified several different configuration files available in the \emph{app} partition of the extracted file system. While some configuration files are related to the behavior of the applications executed by the system at start up (more on this in Section~\ref{ss:bash_script}), by examining certain well-known strings in the text files, we were able to identify two extremely interesting files.

The first file is \texttt{ap\_mode.cfg} (located at the root directory), which contains the default configuration for the access point exposed by the device while in configuration mode. This includes the interface label, the SSID type and prefix, the default IP address (\texttt{192.168.55.1}), DHCP range, and default password in plain text.

The second file is \texttt{shadow} (also located at the root directory), which serves as a modified version of the \texttt{/etc/shadow} file found in all UNIX systems that will be copied in the \texttt{/etc} folder at start up (see Section~\ref{ss:bash_script}). The \emph{shadow} file contains a single user (\texttt{root}) and the hash (generated with \emph{descrypt}) of the password associated with the root user. Since the length of the hash is limited to only $13$ bytes, we attempted to reverse-engineer the password via a brute-force attack, eventually succeeding in recovering it. The recovered password has a length of $8$ characters and is likely generated following a pre-defined scheme used by the vendor. Upon successful authentication with the recovered credentials via UART shell, we were also able to access the U-Boot shell after a couple of easy and intuitive modifications to the password.

\subsection{Service mapping}
\label{ss:service_mapping}
We conducted a network mapping on our device to identify all services accepting incoming traffic from external hosts. This mapping was performed from both an external attacker's perspective (e.g., by mapping the ports from a host connected to the same network as the device) and by accessing the internal shell of the camera (made possible by the recovered password found in Section~\ref{ss:unix_conf}) to identify the services handling incoming traffic.

\subsubsection{External mapping report}
We performed the external mapping using the \emph{nmap} utility~\cite{nmap} to identify all listening ports accepting both TCP and UDP connections. The results of the network mapping procedure indicate that the camera is accepting \emph{telnet} connections on port $23$/TCP, \emph{rtsp} (Real-Time Streaming Protocol) connections on port $8554$/TCP, and exposes other unknown services on ports $843$/TCP, $1300$/TCP, $6688$/TCP, $8699$/TCP, $9876$/TCP, $3702$/UDP, $5012$/UDP, $5683$/UDP, and $19966$/UDP.

We were able to access the camera via the \emph{telnet} protocol using the \texttt{root} credentials recovered in the previous step, and via \emph{rtsp} using the default credentials associated with the service, stored in plain text in the \texttt{ap\_mode.cfg} configuration file. It is worth noting that these credentials are hard-coded and identical in all IP cameras from the same vendor, and cannot be modified via the corresponding smartphone application. Therefore, by accessing a network where a camera from the same vendor as the one used in our analysis is installed, anyone can access its live video stream using the same set of credentials.

\begin{tcolorbox}[width=.98\columnwidth,bicolor,
    left=0mm,right=0mm,top=0mm,bottom=0mm,arc=0mm,
    colback=gray!10!white,colbacklower=red,colframe=black,
    colupper=black,collower=black]
    \textbf{CVE-2023-30351:} Remote access via hard-coded credentials - Tenda IP Camera CP3 was discovered to contain a hard-coded default password for root which is stored using weak encryption. This vulnerability allows attackers to connect to the TELNET service (or UART) by using the exposed credentials.
    \tcblower
    Base Score: \textbf{7.5 High} \\
    \scriptsize{Vector: CVSS:3.1/AV:N/AC:L/PR:N/UI:N/S:U/C:H/I:N/A:N}
\end{tcolorbox}

\begin{tcolorbox}[width=.98\columnwidth,bicolor,
    left=0mm,right=0mm,top=0mm,bottom=0mm,arc=0mm,
    colback=gray!10!white,colbacklower=black,colframe=black,
    colupper=black,collower=white]
    \textbf{CVE-2023-30352:} RTSP feed access via hard-coded credentials - Tenda IP Camera CP3 was discovered to contain a hard-coded default password for the RTSP feed.
    \tcblower
    Base Score: \textbf{9.8 Critical} \\
    \scriptsize{Vector: CVSS:3.1/AV:N/AC:L/PR:N/UI:N/S:U/C:H/I:H/A:H}
\end{tcolorbox}

\subsubsection{Internal mapping report}
Following the mapping of services accepting incoming connections from outside the device, we exploited the previously found access via telnet to discover other potentially interesting services running on the device that could lead to further vulnerability findings. Initially, we analyzed the processes active on the device after gaining root access through the exploited telnet service. We found that only slightly more than $50$ processes were active on the device, many of which were observed starting as system processes via the UART serial interface logs.

Upon observing the list of active threads on the device, we discovered that the vast majority were spawned from two different applications, namely \emph{noodle} and \emph{apollo}, which were also found in our copy of the flash memory in the \emph{app} partition.

We then proceeded with the identification of processes associated with the open ports found in the external network mapping. The results of the \texttt{netstat} command highlighted that the \emph{telnet} connection is managed by \emph{inetd} (as expected), while all other connections are managed by either \emph{noodle} or \emph{apollo}. Specifically, \emph{noodle} listens on ports $843$/TCP, $1300$/TCP, and $5012$/UDP, while \emph{apollo} serves all other ports, including $8554$/TCP (RTSP). Since these two binaries serve as the main entry points for any external connection (including communication with the vendor's servers and the vendor's application), we will focus on their analysis in Section~\ref{s:binal}.

\subsection{Bash scripts}
\label{ss:bash_script}
We analyzed the content of the bash scripts found in the \texttt{/etc/init.d} folder, which contains $5$ different scripts:
\begin{itemize}
    \item \textbf{S01udev:} responsible for creating some system folders and run the \texttt{udevd} demon and to start the \emph{udevstart} program;
    \item \textbf{S02init\_rootfs:} responsible for mounting the \emph{data} file system partition;
    \item \textbf{S03network:} responsible for configuring the network interface;
    \item \textbf{S04app:} responsible for mounting the \emph{app} file system, initializing different applications, and starting the \emph{noodles} application;
    \item \textbf{rcS:} responsible for executing all the scripts found in the \texttt{/etc/init.d} folder starting with the \texttt{S[0-9][0-9]} regex expression.
\end{itemize}

Since the \emph{S04app} script is the most interesting initializing script, we decided to analyze it in detail to recreate the entire initialization process of the applications being executed on the device and to find any potential vulnerabilities. The \emph{S04app} script invokes another script (\texttt{chk\_ver.sh}) that is responsible for updating any script found in the \texttt{/usr/bin} folder with newer versions found in the \texttt{/app} folder, if available. While it is possible to exploit this script to overwrite system applications with another file with the same name placed in the \texttt{/app} folder, it is worth noting that obtaining access to the system is necessary to perform this exploit. Given that the only user available on the device is the \emph{root} user, the process of modifying a script is trivial and of little interest in our case study.

The \emph{S04app} then proceeds to invoke three hard-coded scripts (\texttt{patch.sh}, \texttt{sys\_init.sh}, and \texttt{app\_init.sh}) from the \texttt{/app} directory, if available. Unfortunately, in both our device and flash image, we were unable to locate these scripts, as they are most likely related to the installation of a patch downloaded from the vendor's website.

After executing the \emph{noodles} application, the \emph{S04app} script identifies and mounts the SD card on the \texttt{/mnt/sd} mount point. If the SD card is mounted correctly, it then executes another script (\texttt{iu.sh}), which appears to be an update script. Upon further analysis, the \texttt{iu.sh} script copies the content of a file named \emph{Flash.img} found in the root directory of the SD card to a temporary working folder (\emph{/home}) and proceeds by copying its content to replace the entire flash memory. It is worth noting that since the entire system boot process does not check the integrity of the loaded flash image, it is possible to overwrite the system simply by inserting a properly formatted SD card with the malicious firmware image, without the necessity of accessing the device through the network.

\begin{tcolorbox}[width=.98\columnwidth,bicolor,
    left=0mm,right=0mm,top=0mm,bottom=0mm,arc=0mm,
    colback=gray!10!white,colbacklower=red,colframe=black,
    colupper=black,collower=black]
    \textbf{CVE-2023-30356:} Missing support for Integrity Check - Tenda IP Camera CP3 was discovered missing Support for an Integrity Check, allowing attackers to update the device with crafted firmware.
    \tcblower
    Base Score: \textbf{7.5 High}  \\
    \scriptsize{Vector: CVSS:3.1/AV:N/AC:L/PR:N/UI:N/S:U/C:N/I:H/A:N}
\end{tcolorbox}

\section{Binary analysis: noodles and apollo}
\label{s:binal}
In this section, we present a detailed analysis of the two main binaries responsible for managing network connections on the Tenda CP3 camera, namely \emph{noodles} and \emph{apollo}. The primary objective of the analysis presented in this section is to outline the methodology we adopted in identifying vulnerabilities that could potentially allow for remote code execution on the device.

Our analysis is based on the reverse-engineered representation of the two binaries obtained with Ghidra~\cite{ghidra}, upon which we applied a customized version of the \emph{rhabdomancer} script~\cite{rhabdomancer} to identify insecure functions handling external connections (\texttt{recv}, \texttt{recvfrom}, and \texttt{recvmsg}). Rhabdomancer is a Ghidra script designed to assist with vulnerability research tasks based on a candidate point strategy against software written in C/C++. It locates all calls to potentially insecure functions (the candidate points), which can be used to find insecure input access to the process. Additionally, we developed a tool on top of rhabdomancer that automates the entire process of reconstructing the function call sequence from the \texttt{main} function to the specified entry points~\footnote[2]{A link to the GitHub repository will be included in the camera-ready version of the manuscript}, enabling us to easily map all the threads of the two processes to the different ports they use. Subsequently, we proceeded with a manual analysis to identify potential security vulnerabilities and design exploits to achieve remote code execution on our device.

\subsection{The \emph{noodles} binary}
\label{ss:noodles}
By executing our modified version of the rhabdomancer script, which specifically targeted functions relevant to our analysis (primarily \emph{recv}, \emph{recvfrom}, and \emph{recvmsg}), we identified three candidate points in the decompiled binary. These points were labeled as \texttt{FUN\_00014e68}, \texttt{FUN\_0001fc14}, and \texttt{FUN\_00012b7c}.

\subsubsection{\texttt{FUN\_00014e68}}
This function is referenced in $5$ other functions within the \emph{noodles} binary. We will refer to these different invocations using the memory addresses of our reversed binary for simplicity.

\textbf{1. 0x00011b04} The first reference is inside the \texttt{main} function, and is related to a socket listening on port $1300$ (one of the ports already identified in the previous analysis step). In this invocation, \texttt{FUN\_00014e68} is used to receive commands from the client, as confirmed by the string \texttt{receive\_cmd from client<\%d>: <\%s> len = \%d \textbackslash n} found a couple of instructions later. The available commands (hard-coded in the binary) through this interface include some suspicious strings like \texttt{ELFEXEC}, \texttt{DOWNLOAD}, and \texttt{SYSTEM}.

\textbf{2. 0x000123bc} The second reference to our target function is inside the \texttt{FUN\_00012110} function, which is referenced by $3$ other functions: \texttt{FUN\_000128a0} (invoked by the \texttt{main} function after the \texttt{ELFEXEC} command is received), \texttt{FUN\_00014674} (invoked by the \texttt{main} function after the \texttt{DOWNLOAD} command is received), and \texttt{FUN\_000147ac} (invoked by the \texttt{main} function after the \texttt{UPGRADE} command is received). Upon further investigation, we confirmed that these references are all related to the previously discussed functionalities, and they can be used to trigger the execution of different scripts available on the camera.

\textbf{3. 0x0001272c} The third reference is found inside two different functions: \texttt{FUN\_000146e4} (invoked by the \texttt{main} function after the \texttt{UPLOAD} command is received) and \texttt{FUN\_00014748} (invoked by the \texttt{main} function after the \texttt{FLASHDUMP} command is received). While via the former invocation it is possible to upload a specific file to the camera, by exploiting the latter invocation we demonstrated that it is possible to remotely upload a modified version of the firmware that will be copied to the flash memory on the next restart. This enables the exploitation of the vulnerability identified in Section~\ref{ss:bash_script} (\textbf{CVE-2023-30356}) without requiring physical access to the camera to upload a modified firmware version on the SD card.

\textbf{4. 0x00013cbc} The fourth reference is found inside the function \texttt{FUN\_00013c30}, which happens to be the \emph{policy\_thread} spawned by the function \texttt{FUN\_00013df4} (directly called in the \texttt{main} function). The \emph{policy\_thread} is listening on port $843$, accepts a fixed string (\texttt{policy-file-request}), and responds with a fixed XML structure.

\textbf{5. 0x000144f8} The final reference is found inside \texttt{FUN\_000143c0}, which is another function associated with the management of external commands accepted by \emph{noodles}. Specifically, this function is called upon the reception of the \texttt{SYSTEMEX} command, which has already been exploited by other researchers in \textbf{CVE-2023-23080} to achieve remote code execution.

\subsubsection{\texttt{FUN\_0001fc14}} 
The second function containing a \texttt{recv} invocation is referenced only once in function \texttt{FUN\_0001d2c8} in the \emph{noodles} binary. However, this latter function is referenced three times.

The first reference (\texttt{FUN\_0001d2c8}) is related to some WiFi connection tests and extends up to the \texttt{main} function (invoked after receiving a \texttt{SYSTEM} command containing the \emph{STATUS} keyword). We analyzed the entire activation graph of this invocation, composed of $6$ different functions, and established that this set of functions is used to test the status of a known WiFi SSID by the camera on system boot.

The second reference (\texttt{FUN\_0001d1f8}) is also related to WiFi communication and is also invoked indirectly by the \texttt{main} function after a \texttt{SYSTEM} command is received. However, this second function accepts a different system command (\emph{SCAN}) and performs a network scan on the \texttt{wlan0} (hard-coded) network interface.

The third reference (\texttt{FUN\_0001d3b0}) is once again related to WiFi communication and invoked via the \texttt{SYSTEM} handler of the \texttt{main} function. This time, the function accepts a \emph{SCAN\_RESULTS} command and prints the output of the last saved network scan on the serial interface.
We remark that all three of these handlers return the same value (\texttt{<SYSTEMEX\_ACK>ok</SYSTEMEX\_ACK>}), which is the default response for any successful connection via the \texttt{SYSTEM} command.
Additionally, we note that all of these commands are vulnerable to remote code injection, and that any received command is executed on the device with root privileges without requiring authentication.

\subsubsection{\texttt{FUN\_00012b7c}} 
The third function containing a \texttt{recvfrom} function is actually the \emph{multicast\_thread} spawned by the \texttt{main} function and attached to port $5012$/UDP. The \emph{multicast\_thread} is configured to accept two commands, namely \texttt{YGMP\_SVR} and \texttt{YGMP\_CMD}.

Upon reception of the former command, \emph{noodles} opens different configuration files to read the current settings of the device, which are then returned as an XML structure to the caller. These settings include the \textit{IP} and \textit{MAC} addresses of the camera, its \textit{serial number}, the value encoded in the \textit{QR Code}, the \textit{hardware version}, and other information.

The latter command (\texttt{YGMP\_CMD}), however, is far more interesting in the scope of our work, as it allows unauthenticated remote code execution on the camera by sending a formatted XML payload. In particular, the payload accepts $3$ different tags for parsing: \emph{TARGET}, \emph{MAC}, and \emph{CMD}. Although the content of both \emph{TARGET} and \emph{MAC} are apparently not used except in some printing functions, the content of the \emph{CMD} tag is compared to the \emph{reboot} string. If the \texttt{strcmp} returns $0$, the \texttt{FUN\_00016ea8} function is called with the argument \texttt{/app/bin/cmd reset}; otherwise, the content of the \emph{CMD} tag is passed directly to the same function.

Upon further inspection, we verified that the \texttt{FUN\_00016ea8} function is a simple wrapper for the \texttt{system} function, with the arguments passed to the function being forwarded directly to \texttt{system} without proper sanitization. This allows unauthenticated remote code execution on the camera by simply passing a command different from \emph{reboot}.

\begin{tcolorbox}[width=.98\columnwidth,bicolor,
    left=0mm,right=0mm,top=0mm,bottom=0mm,arc=0mm,
    colback=gray!10!white,colbacklower=black,colframe=black,
    colupper=black,collower=white]
    \textbf{CVE-2023-30353:} Unauthenticated RCE - Tenda IP Camera CP3 allows unauthenticated remote code execution via an XML document.
    \tcblower
    Base Score: \textbf{9.8 Critical} \\
    \scriptsize{Vector: CVSS:3.1/AV:N/AC:L/PR:N/UI:N/S:U/C:H/I:H/A:H}
\end{tcolorbox}

We emphasize that by following the described methodology, which is based on the analysis of the \texttt{recv} functions and their activation path, we were able to find all handlers for incoming connections managed by the \emph{noodles} application. Furthermore, we demonstrated how to exploit these handlers to obtain unauthenticated RCE with root privileges on the device.

\subsection{The \emph{apollo} binary}
\label{ss:apollo}
We employed the same methodology applied on the \emph{noodles} binary to analyze the \emph{apollo} binary. In particular, we want to remark that the vanilla rhabdomancer script on the whole \emph{apollo} binary resulted in more than $128000$ candidate points, while our tool returned less than $100$ points, which we further reduced with a simple duplicate removal of different points on the same function calls. 

The \emph{apollo} binary utilizes $65$ different threads to perform various tasks based on commands received from $7$ different ports. Each of these threads eventually leads to one or more of the $25$ \texttt{recv} functions. These functions containing the \texttt{recv} calls are solely responsible for managing incoming data, while the parsing of the received data structure is handled by the calling functions. 

After thorough analysis, we successfully mapped each thread to a specific port listened to by the \emph{apollo} process and identified the exposed functionalities.

\begin{itemize}
    \item \textbf{3702/TCP} exposes ONVIF~\cite{onvif} \emph{discovery}, \emph{notification} and \emph{hello} threads as required by the ONVIF Core Specification [$3$ threads];
    \item \textbf{6688/TCP:} exposes an HTTP server [$3$ threads];
    \item \textbf{8554/TCP:} exposes the \emph{RTSP} service, which is used to access the camera video and audio [$2$ threads]
    \item \textbf{8699/TCP:} exposes a set of threads related to manage different functionalities of the camera [$52$ threads];
    \item \textbf{9876/TCP:} exposes the \emph{yserver} TCP handler for incoming connections [$3$ threads];
    \item \textbf{5683/UDP:} exposes the \emph{COAP} (COnstrained Application Protocol~\cite{coap}) functionalities [$1$ thread];
    \item \textbf{19966/UDP:} exposes the \emph{yserver} UDP handler for incoming connections [$1$ thread].
\end{itemize}

We proceeded with a more in-depth analysis of the main handler of port \textbf{8699/TCP} (\texttt{ut\_cmd\_server\_init}) due to the significant number of threads directly associated with this port. We discovered that this connection is utilized for direct communication with the user application within the same network or via cloud services when in a different network. It accepts all possible commands that the user can provide through the application interface.

The incoming packets are handled by a dedicated thread (\texttt{ut\_rcmd\_server\_proc}) and then passed to a parser function to identify the received command (\texttt{FUN\_0007cb00}). In this function, the parser only checks that the received command starts with the character \texttt{!} and then proceeds to compare the remainder of the received string with hard-coded commands. If the received string matches one of the commands, then a thread is spawned and the desired function is executed on the camera. We identified $148$ different commands ($147$ plus the \texttt{help} command, which returns the description of all the other commands), which are mapped to $50$ different threads. As an example, the \texttt{audio\_output\_proc} manages various functionalities related to recording audio from the microphone (e.g., \emph{audio\_vol\_in}, used to set the audio input volume, or \emph{capture\_audio}, used to capture audio from the microphone) and setting different speaker parameters. Meanwhile, functionalities related to playing audio on the camera are managed by the \texttt{audio\_output\_proc} (e.g., \emph{loop\_audio} and \emph{play\_audio}).
However, if the string received by the \texttt{ut\_rcmd\_server\_proc} thread and parsed by the \texttt{FUN\_0007cb00} function is not recognized as one of the commands directly managed by the process, the \texttt{system} function is executed by passing the remaining string as the only parameter. By supplying a correctly formatted string to the service, we were able to achieve another instance of unauthenticated remote code execution on the camera.

Finally, we note that we also identified potential exploits on the other ports handled by the \emph{apollo} process. While the methodology presented in this paper has proven effective in identifying vulnerabilities on our device, we emphasize that the process of behavior reconstruction (i.e., the detailed analysis presented on the \emph{noodles} binary) and the subsequent crafting of exploits to verify a security vulnerability still heavily rely on human skill and experience.

\section{Conclusions}
\label{s:conclusions}
This paper proposes a novel approach based on classical reverse engineering methodologies for in-depth security analysis of consumer IP cameras. The methodology comprises five steps:
\begin{enumerate}
\item gathering relevant information from open sources; 
\item physical access to a IP camera specimen aiming at identifying low-level attack vectors;
\item firmware extraction and static analysis of the whole file system, including configuration files, scripts and executables;
\item dynamic analysis of the network behavior, aiming at identifying all remote attack surfaces (such as open TCP and UDP port) of the connected IP camera;
\item in-depth reversing of all executables that implement network-facing services.
\end{enumerate}

We provide a detailed example of the application of the proposed methodology by taking the widespread Tenda CP3 IP camera as a relevant use case. Our methodology allowed us to identify five novel CVEs with a CVSS score ranging from $7.5$ to $9.8$. 

We remark that the proposed methodology differs from the approaches that are commonly proposed in many related works, which only perform network-based analysis and fall short from executing a complete reversing of relevant executables. 

To partially automate our approach we also developed a novel tool~\footnote[4]{A link to the GitHub repository will be included in the camera-ready version of the manuscript} based on Ghidra and rhabdomancer that is able to identify the few functions that manage data received from network connections within large binary executables.

\section*{Responsible Disclosure}
Before publication, we contacted Shenzen Tenda Technology Co., Ltd. in May $2023$ and disclosed our initial findings to them. We informed their representative that we had discovered several vulnerabilities in one of their products and mutually agreed to proceed with the responsible disclosure procedure once the CVE IDs for the vulnerabilities were assigned. We obtained $5$ CVE IDs (out of the $7$ initially requested, with $2$ CVE IDs covering the missing ones) at the beginning of June $2023$, and promptly reached out to their representative. As of today, we have not received any response from their representative despite multiple attempts to contact them. We have decided to publicly release the results of our work only now, following the end of the embargo in August $2023$, in the hope of receiving a response from the Tenda representative in the meantime. We believe that the severity of the vulnerabilities found in their product is critical and requires immediate attention. We will continue to reach out to the company's representative with the hope that the identified vulnerabilities are acknowledged by the company and addressed in the best possible manner.

\section*{Acknowledgments}
This work was partially supported by project SERICS (PE00000014) under the MUR National Recovery and Resilience Plan funded by the European Union - NextGenerationEU.

\small{
\bibliographystyle{IEEEtran}
\bibliography{bibliography}{}
}


\end{document}